\documentstyle[12pt]{article}
\newcommand{\be}{\begin{equation}}
\newcommand{\ee}{\end{equation}}
\newcommand{\bea}{\begin{eqnarray}}
\newcommand{\eea}{\end{eqnarray}}

\newcommand{\fa}{$\phi_{\alpha}(x)$}
\newcommand{\fas}{$\phi_{\alpha}^{+}(x)$}

\newcommand{\Aa}{$A_{\alpha}({\bf k})$}
\newcommand{\Aas}{$A_{\alpha}^{+}({\bf k})$}
\newcommand{\Abs}{$A_{\beta}^{+}({\bf q})$}

\newcommand{\vac}{\left.\mid 0\right>}
\newcommand{\k}{{\bf k}}
\newcommand{\q}{{\bf q}}
\newcommand{\x}{{\bf x}}
\newcommand{\y}{{\bf y}}
\newcommand{\r}{{\bf r}}
\newcommand{\R}{{\bf R}}
\newcommand{\p}{{\bf P}}
\newcommand{\n}{\bigtriangledown}
\begin{document}
\vspace{8cm}
\begin{flushright}
ISU-IAP.Th94-02, Irkutsk
\end{flushright}

\begin{center}
{\large \bf Solution of The Heisenberg Equation \\
For The Four-fermion Contact Interaction by The Method \\
of Dynamical Mappings.}
\vskip 1cm
{\bf A.N.Vall, V.M.Leviant, A.V.Sinitskaya}
\vskip .5cm
\em {Department of Theoretical Physics,\\
 Irkutsk State University, \\
 Irkutsk, RUSSIA.\\
e-mail: VALL@physdep.irkutsk.su}
\end{center}

\begin{abstract}
In the paper we consider non-relativistic analog of the Nambu-Jona-Lasinio
model and demonstrate, by this most simplified example, the method of
dynamical mappings of Heisenberg fields on "physical" fields. We obtain
the expressions for one particle eigen-excitation spectrum of full
Hamiltonian (spectrum of "clad" fermions), wave function of two
excitations' bound state and its mass, and compute the Bethe-Solpither
amplitude. We derive as well the relation for the density of vacuum
energy as a function of mapping parameters. The minimization of energy leads
us to the fixation of these parameters. The method admits the generalization
on the relativistic case.
\end{abstract}

\newpage

\begin{center}
{\bf Introduction}
\end{center}
\vskip 2mm

The present paper is devoted to the solution of the Heisenberg
equation for four-fermion contact interaction (non-relativistic analog of
the Nambu-Jona-Lasinio (NJL) model without form factor\cite{NJL,Ber}) in
representation of physical particles and their states. The idea of "physical"
states was first introduced by Heisenberg\cite{Heis}. However, the most
rigorous and consistent realization of the physical states is given in
the papers\cite{Umez}. The Hamiltonian of the interacting particles system
on these states by definition has the diagonal form:
\be
\left< a\mid H\mid b\right> = < a\mid \int d^3q A^+(\q)A(\q)
E(\q) \mid b> + W_0<a\mid b>. \label{I1}
\ee
The important fact is that the states $A^+(\k)\vac$ are eigenstates of the
Hamiltonian $H$, that is
\bea
\left[H,A^+(\k)\right]\vac &=& E(\k)A^+(\k)\vac  \label{I2}  \\
A(\k)\vac &=& 0, \nonumber
\eea
where $E(\k)$ determines thus the spectrum of one particle excitations of
"physical" vacuum, defined by the second line in (\ref{I2}).

Since all interaction characteristics are reduced to the calculating of
matrix elements of Heisenberg field combinations over the "physical" vacuum,
it is necessery to have a connection between the Heisenberg fields and
the "physical" ones, whose excitations are described by the $A^+(\k)$
operators. We will call such a connection, following the papers\cite{Umez},
the dynamical mapping of Heisenberg fields on "physical" fields. The regular
procedure of seeking the dynamical mappings is based on the fact
that they must be consistent both with the Heisenberg equation for
interacting fields and with relation (\ref{I2}). The meaning of the latters,
in fact, is led to the statement that the state $A^+(\k)\vac$ describes free
quasiparticle with the momentum $\k$ and the energy $E(\k)$. It is worth
noting that the mehtod of the dynamical mappings is naturally adjusted to
describing the phase transition phenomena and the processes with the
spontaneous symmetry breaking. This fact is connected with the known case,
when the dynamical mappings correspond to the unitary non-equivalent
transformations of the Hilbert space, transfering it to the space ortogonal
to the initial one, accompanied by the reconstruction of the Hamiltonian
spectrum and, in particular, by changing the vacuum (ground state)
energy. We show below that the vacuum energy density depends on the
parameters contained in the dynamical mapping. By varying the values of these
parameters we "enumerate" all possible mutually ortogonal Hilbert spaces. One
of these spaces is the "physical" space, and the criterion of its choosing is
the condition for the vacuum state energy to be minimal.

Further on we consider the bound state of two excitations (of two "clad"
fermions). It is well known, that while investigating the bound states
properties in quantum field theory the main quantity is the Bethe-Solpither
(BS) amplitude\cite{BS}, given by the following matrix element:
\be
\Phi_{abc}(x,y;\k) = \left< 0\mid T A_a(x)B_b(y)\mid A_c(\k)
\right>, \label{I3}
\ee
where $A_a(x)$ and $B_b(y)$ are field operators in the Heisenberg
representation, and $\left.\mid A_c(\k)\right>$ is the bound state vector,
$a,b,c$ - generalized indices (spin, flavour, colour, {\it ect}), depending
on the choice of studied theory. The usual approach is to use the BS
equation\cite{BS} ether on $\Phi_{abc}(x,y;\k)$ or on vertex function:
\be
\Gamma_{ab\dots}(x_1,x_2,\dots) = \left< 0\mid T A_a(x_1)B_b(x_2)\dots
\mid 0\right>. \label{I4}
\ee
The type of the vertex is fixed by the interaction. It is worth mentioning,
that reactions of composite particles can be described by the effective
Lagrangians (see, for example, reviews\cite{Volk, Leut}). Development of QCD
initiated the interest to the models with four-fermion
interaction\cite{Perv}(non-local prototypes of the NJL model) in which gluons
are effectively accounted by introducing the non-local quark interaction
kernel\cite{Perv}.

The equations on the function (\ref{I3}) or (\ref{I4}) are usually solved by
using some perturbative theory (loop expansion, "ladder" approximation,
{\it ect}). This leaves the set of crucial points out of consideration.
Firstly, the matter of vacuum state choice. Secondly, connection of
initial field operators in Heisenberg representation with operator
$\hat{A}^+(\k)$, forming the bound state. At last, for the solution of BS
equation it is necessery to know total two-point Green functions, i.e. the
solution of the Schwinger-Dyson (SD) equation. The method of dynamical
mappings allows to overcome all these dificulties, because it gives the
opportunity to do direct calculations of all interaction physical
characteristics.
\vspace{1cm}

{\bf 1. Dynamical mapping of the Heisenberg fields}
\vskip 2mm

The Hamiltonian of the four-fermion interaction can be written in the
following form:
\be
H = \int d^{3}x \left[\psi_\alpha^{+}(x)\epsilon(\n)\psi_\alpha(x) +
\frac{\lambda}{4} \chi^{+}(x)\chi(x) \right],
\label{1}
\ee
where $\alpha$ is spin index running over $1,2, \epsilon (\n)$ -
energy spectrum of free fermions, defined by the condition
\be
\epsilon(\n)e^{i{\k \x}} = \epsilon({\k})e^{i{\bf kx}}
\label{2}
\ee
and
$$
\chi(x) = \epsilon_{\alpha \beta}\psi_{\alpha}(x)\psi_{\beta}(x),\;\;
\chi^{+}(x) = \epsilon_{\alpha \beta}\psi^{+}_{\beta}(x)\psi^{+}_{\alpha}(x).
$$
We need to find the relation between the Heisenberg fields
$\psi_{\alpha}(x)$ and fields $\phi_{\alpha}(x)$, whose excitations would
be eigenstates of the Hamiltonian $H$. In other words, it is necessary
to find out the dynamical mapping \cite{Umez}. As a starting point we
consider the following case. Let the fields $\psi_{\alpha}(x)$ satisfy the
canonical equal-time anticommutation relations:
\be
\left\{\psi^{+}_{\alpha}(x),\psi_{\beta}(y)\right\}_{t_{x}=t_{y}} =
\delta_{\alpha \beta}\delta(x - y).
\label{3}
\ee
Defining generator $W$ as:
\be
W = \int d^{3}x\left(f(x)\chi(x) - \overline{f}(x)\chi^{+}(x)\right),
\label{4}
\ee
where $f(x)$ is arbitrary complex function, we can write down the
transformation of the fields $\psi_{\alpha}(x) \rightarrow
\phi_{\alpha}(x)$
\be
\phi_{\alpha}(x) = e^{W}\psi_{\alpha}(x)e^{-W}.
\label{5}
\ee
In this equation all the fields are
taken at equal times. Using equation (\ref{3}), we obtain:
\bea
\phi_{\alpha}(x) = u(x)\psi_{\alpha}(x) - v(x)\epsilon_{\alpha \beta}
\psi^+_{\beta}(x)  \nonumber \\
\psi_{\alpha}(x) = u(x)\phi_{\alpha}(x) + v(x)\epsilon_{\alpha
\beta}\phi^+_{\beta}(x)
\label{6}
\eea
where
$$
u(x) = cos\mid 2f(x)\mid,\;\;\;\;\; v(x) = \frac{\overline{f}(x)}{f(x)}
sin\mid 2f(x)\mid
$$
 From the explicit form of $u(x)$ and $v(x)$
there follows the normalization condition:
\be
u^{2}(x) +\mid v^{2}(x)\mid = 1
\label{7}
\ee
Second equation in (\ref{6}) gives rise to
the dynamical mapping of the Heisenberg fields $\psi_{\alpha}(x)$ on the
physical fields $\phi_{\alpha}(x)$ if, certainly, it is consistent with
the Hamiltonian (\ref{1}). By the next step we postulate the set of
properties for the field $\phi_{\alpha}(x)$. That is, let
\be
\phi_{\alpha} (x) = \frac{1}{(2\pi)^\frac{3}{2}}\int d^{3}k g({\k})
e^{i\k \x - iE(\k)t}A_{\alpha}(\k).
\label{8}
\ee
be a representation for \fa, where
\Aa and \Aas are annihilation and creation operators for the particles with
momentum ${\bf k}$, satisfying the following canonical relations:

\be
\left\{A_{\alpha}(\k), A^{+}_{\beta}({\bf q})\right\} =
\delta_{\alpha \beta}\delta (\k - {\bf q}).
\label{9}
\ee
$E(\k)$ is unknown yet excitation spectrum of \Aas;
$g(\k)$  is a distribution function over the momenta  inside  the packet,
and $g(-\k) = \overline{g}(\k)$.  Here  we should note the following
important circumstance.  Representation (\ref{8}) completely  breaks  the
initial properties of transformations  (\ref{5}) and (\ref{6})  - they
cease  to be canonical, i.e. they do  not preserve the  anticommutation
relations. It  is the presence  of  $g(\k)$  function  that  causes  the
field   anticommutation relations of  \fa not  to be  local. Indeed,  from
equation  (\ref{8}) with account of relation (\ref{9}) we have:

\bea
&&\left\{\phi_{\alpha}(x), \phi^{+}_{\beta}(y)\right\} =
\frac{\delta_{\alpha \beta}}{(2\pi)^{3}}\int d^{3}k\mid g(\k)\mid^{2}
e^{[i\k(\x -{\bf y}) - iE(\k)(t_{x} - t_{y})]} \nonumber \\
&&\left\{\phi_{\alpha}(x), \phi^{+}_{\beta}(y)\right\}_{t_{x} = t_{y}} =
\frac{\delta_{\alpha \beta}}{(2\pi)^{3}}\int d^{3}k\mid g(\k)\mid^{2}
e^{i\k(\x -{\bf y})} \nonumber \\
&&\left\{\phi_{\alpha}(x), \phi^{+}_{\beta}(y)\right\}_{x = y} =
\frac{\delta_{\alpha \beta}}{V^{*}} \label{10} \\
&&\left\{\phi_{\alpha}(x), A^{+}_{\beta}(\k)\right\} = \frac{\delta_{\alpha
\beta}}{(2\pi)^{\frac{3}{2}}}g(\k)e^{i\k \x}. \nonumber
\eea
where
\be
\frac{1}{V^{*}} = \frac{1}{(2\pi)^{3}}\int d^{3} k\mid g(\k)\mid^{2}
\label{11}
\ee
 So, the input of $g(\k)$ allows to modify the
product of fields to be defined  at one point, thus breaking the  locality.
Transformation  (\ref{5})  plays  auxiliary  role  for  obtaining
(\ref{6}) leading to the fact  that relations (\ref{8}) and (\ref{9})  have
no sense any longer.

To find the meaning of normalization (\ref{11}) we can use the dimensions of
the fields (as it follows from (\ref{9})) $[A_{\alpha}^{2}(\k)] = L^{3},
[\psi_{\alpha}^{2}(x)] = L^{-3}$, therefore, $g(\k)$ is dimensionless
quantity, and integral in the l.h.s of (\ref{11}) has the dimension $L^{-3}$.
Further on, since $\mid g(\k)\mid^{2}$ characterizes the momentum
fluctuations inside the excitation, volume $V^{*}$ has to be connected
with a characteristic spatial size of the excitation. If one accepts that
in $(\Delta k)^{3}$ region, where $\mid g(\k)\mid^{2}$ essentially differs
from zero, it is equal to $1$ ($\mid g(\k)\mid^{2}\approx  1$), then
\be
\frac{1}{(2\pi)^{3}}\int d^{3} k \mid g(\k)\mid^{2} = \frac{1}{(2\pi)^{3}}
(\Delta k)^{3} = \frac{1}{L^{*3}}
\label{12}
\ee
where $L^{*} = {{2\pi}\over{\Delta k}}$ is linear size of the packet.
Accounting this fact we take for $\mid g(\k)\mid^{2}$ the normalization
(\ref{11}) with the additional condition:
\be
<\mid g(\k)\mid^{2}>_{g} = {\int d^{3}k \mid g(\k)\mid^{4}\over
\int d^{3}k \mid g(\k)\mid^{2}} = 1.
\label{13}
\ee
Thus, in momentum space typical for the wave packet \fas,
$\mid g(\k)\mid^{2}\sim 1$.

So, we would search a solution in the form of relation (\ref{6}) with
conditions (\ref{8}), (\ref{10}), (\ref{11}) and(\ref{13}). Let us
integrate the left and right parts of the equation (\ref{6}) with the plain
wave and obtain expressions of Heisenberg operators $a_\alpha (\bf k)$ via
creation \Aas and annihilation \Aa operators of quasi-particle:
\be
a_{\alpha}(\k) = \int d^{3} k'g(\k')u(\k - \k')A_{\alpha}(\k') +
\int d^{3} k'g(\k')v(\k - \k')\epsilon_{\alpha \beta}A^{+}_{\beta}(\k').
\label{14}
\ee
Hence we can see that transformations (\ref{6}), in contrast to the
Bogolubov transformations, are nonlocal over momentum. Therefore, vacuum
state $\mid 0 >$ defined as \Aa$\mid 0> = 0$, should essentially differ
from the ground state defined by the Bogolubov transformations. The
question is, whether the transformations (\ref{6}) are "compatible"
(consistent) with the coupling dynamics of the Hamiltonian (\ref{1})? It
will be shown later on that at a particular choice of $u(x)$ and $v(x)$
such a compatibility takes place.

\vskip 1cm
{\bf 2. One-particle excitation spectrum}
\vskip 2mm

Let us choose the parameters $u(x)$ and $v(x)$ as follows
\bea
u(x) = u_{0} = const, & &v(x) = v_{0}e^{i\alpha(x)},
v_{0} = const \nonumber \\
u_{0}^{2} + v_{0}^{2}&=&1  \label{15}
\eea
Then  the dynamical mapping will be of the form:
\bea
\psi_{\alpha}(x) &=& u_{0}\phi_{\alpha}(x) + v_{0}e^{i\alpha (x)}\epsilon_{
\alpha \beta}\phi^{+}_{\beta}(x)  \nonumber \\
\psi ^{+}_{\alpha}(x) &=& u_{0}\phi^{+}_{\alpha}(x) + v_{0}e^{-i\alpha (x)}
\epsilon_{\alpha \beta}\phi_{\beta}(x)  \label{16}
\eea
Our aim is to make the state \Aas  $\vac$ to be an eigenstate of the
Hamiltonian (\ref{1}), with the vacuum being defined with respect to the
operator \Aa
\be
A_{\alpha}(\k)\vac = 0
\label{17}
\ee
Acting by the Hamiltonian (\ref{1}) to the one-particle excitation we have:
\be
H A_{\alpha}^{+}(\k) \vac = A_{\alpha}^{+}(\k) H\vac + \left[H,
A_{\alpha}^{+}(\k)\right]\vac .
\label{18}
\ee
If the vacuum is the Hamiltonian's eigenstate, then for the one-particle
excitation to be stationary it is necessary to satisfy the condition:
\be
\left[ H, A_{\alpha}^{+}(\k)\right] \vac = E(\k) A_{\alpha}^{+}(\k) \vac.
\label{19}
\ee
where $E(\k)$ is the energy of this excitation above the vacuum $\vac$.
This relation can be thought either as a definition of the excitation
spectrum or as the Heisenberg's stationary equation.

Leaving for the moment aside the problem of stationarity of the vacuum,
let us consider only the equation (\ref{19}). The commutator in the l.h.s.
of (\ref{19}) splits to the commutators with the kinetic term and the
coupling term of the Hamiltonian independently, leading to the relations:
\bea
&&\left[
\chi^{+}(x)\chi (x), A^{+}_{\alpha}(\k)\right] = \chi^{+}(x)\left[ \chi (x),
A^{+}_{\alpha}(\k)\right] - \left[ A^{+}_{\alpha}(\k), \chi^{+}(x) \right]
\chi(x), \nonumber \\
&&\left[ \chi (x), A^{+}_{\alpha}(\k)\right] =
\frac{2u_{0}}{(2\pi)^{ \frac{3}{2}}} g(\k) e^{i\k \x}\epsilon_{\beta
\alpha}\psi_{\beta}(x) \label{20} \\
&&\left[ A^{+}_{\alpha}(\k),
\chi^{+}(x)\right] = - \frac{ 2\overline{v}(x)}{(2\pi)^{\frac{3}{2}}} g(\k)
e^{i\k \x} \psi^{+}_{\alpha}(x). \nonumber
\eea

To derive (\ref{20}) we have used the dynamical mapping (\ref{16}).
Putting together all three relations of (\ref{20}) we obtain:
\bea
\left[ \chi^{+}(x) \chi (x), A^{+}_{\alpha}(\k)\right]&=&\frac{2u_{0}}
{(2\pi)^{\frac{3}{2}}} g(\k) e^{i\k \x}\chi^{+}(x) \epsilon_{\beta \alpha}
\psi_{\beta}(x) + \nonumber \\
&+& \frac{2\overline{v}(x)}{(2\pi)^{\frac{3}{2}}} g(\k) e^{i\k \x}
\psi^{+}_{\alpha}(x) \chi (x). \label{21}
\eea
So, the problem is now reduced to the calculation of two terms in the r.h.s.
of (\ref{21}). For the first one we get the following:
\bea
&&\chi^{+}(x)\psi_{\gamma}(x)\vac = v(x)\epsilon_{\gamma \gamma '}\chi^{+}(x)
\phi_{\gamma '}^{+}(x) \vac = \nonumber \\
&& = v(x)\epsilon_{\gamma \gamma '}\epsilon_{
\alpha \beta}\psi^{+}_{\beta}(x)\psi^{+}_{\alpha}(x)\phi_{\gamma '}^{+}(x)
\vac = \nonumber \\
&& = v(x)\epsilon_{\gamma \gamma '}\epsilon_{\alpha \beta}\left( u_{0}
\phi_{\beta}^{+}(x) + \overline{v}(x)\epsilon_{\beta \beta '}\phi_{\beta'}(x)
\right)\left( u_{0}\phi_{\alpha}^{+}(x) + \right.\nonumber \\
&& \left.+ \overline{v}(x)\epsilon_{\alpha
\alpha '}\phi_{\alpha '}(x)\right)\phi_{\gamma '}^{+}(x)\vac = \nonumber \\
&& = v(x)\epsilon_{\gamma \gamma '}\epsilon_{\alpha \beta}\left( u_{0}
\overline{v}(x)\epsilon_{\alpha \alpha '}\phi_{\beta}^{+}(x)\left\{
\phi_{\alpha '}(x), \phi_{\gamma '}^{+}(x)\right\} \right. + \nonumber \\
&& + u_{0}\overline{v}(x)\epsilon_{\beta \beta
'}\phi_{\beta'}(x)\phi_{\alpha}^{+}(x) \left.
\phi_{\gamma '}^{+}(x)\right)\vac = \nonumber \\
&&=\frac{u_{0}v_{0}^2}{V^{*}}\epsilon_{\gamma \gamma '}\epsilon_{ \alpha
\beta}\left( \epsilon_{\alpha \gamma '}\phi_{\beta}^{+}(x) - \epsilon_{\beta
\gamma '}\phi_{\alpha}^{+}(x) + \epsilon_{\beta \alpha}\phi_{\gamma
'}^{+}(x)\right) \vac = 0, \label{22}
\eea
since the last expression in
brackets is fully antisymmetric by indices $\alpha \beta \gamma$.  Consider
now the second term of (\ref{21}).
\bea
&&\psi^{+}_{\alpha}\chi (x)\vac=\epsilon_{\beta \beta '}\psi_{\alpha}^{+}
(x)\psi_{\beta}(x)\psi_{\beta '}(x)\vac= \nonumber \\
&&=-v(x)\psi_{\alpha}^{+}(x)\psi_{\beta}(x)\phi_{\beta}^{+}(x)\vac=
- v(x) \left( u_{0}\phi^{+}_{\alpha}(x) + \bar{v}(x)
\epsilon_{\alpha \alpha '}\phi_{\alpha '}(x)\right)\times \nonumber \\
&&\times\left( u_{0}\phi_{\beta}(x)\phi^+_{\beta}(x)
+ v(x)\epsilon_{\beta \beta '}\phi^{+}_{\beta '}(x)\phi^{+}_{\beta}(x)
\right)\vac= \nonumber \\
&&= - v(x)\left( 2\frac{u_{0}^{2}}{V^{*}}\phi^{+}_{\alpha}(x)\right) - \left(
v_{0}^{2}\epsilon_{\alpha \alpha '}\epsilon_{\beta \beta '}\phi^{+}_{
\beta '}(x)\phi_{\alpha '}(x) \phi^{+}_{\beta} (x)\right)\vac = \nonumber \\
&&=- \frac{2v(x)}{V^{*}}\phi^{+}_{\alpha}(x)\vac.
\label{23}
\eea
Here we have used the constraint $u_{0}^{2} + v_{0}^{2} = 1$
With account of the results (\ref{22}), (\ref{23}) and the expression
(\ref{21}) we will have:
\be
\left[ \frac{\lambda}{4}\chi^{+}(x)\chi (x),
A^{+}_{\alpha}(\k)\right]\vac = - \lambda
v_{0}^{2}\frac{g(\k)}{V^{*}(2\pi)^{\frac{3}{2}}}e^{i\k \x}
\phi^{+}_{\alpha}(x)\vac. \label{24}
\ee
Integrating this relation over the whole configuration space the
contribution into the energy spectrum of the Hamiltonian reads
\be
\left[\frac{\lambda}{4}\int d^3 x \chi^+ (x)\chi (x), A^+_\alpha (\k)\right]
\vac = - \lambda\frac{v_o^2}{V^*}\mid g(\k)\mid^2 A^+_\alpha (\k)\vac
\label{25}
\ee
Now we derive the kinetic term contribution:
\bea
&&\left[\psi^+_{\alpha '}(x)\epsilon (\n)\psi_{\alpha '}(x),
A^+_\alpha(\k)\right] \vac=-\psi^+_{\alpha '}(x)A^+_\alpha (\k)\epsilon (\n)
\psi_{\alpha '}(x)\vac+ \nonumber \\
&&\psi^+_{\alpha '}(x)\epsilon (\n)\left\{ \psi_{\alpha '}(x),A^+_\alpha
(\k)\right\}\vac -
A^+_\alpha (\k)\psi^+_{\alpha '}(x)\epsilon (\n)\psi_{
\alpha '}(x)\vac= \nonumber \\
&&=-\left\{ \psi^+_{\alpha '}(x), A^+_\alpha (\k)\right\}\epsilon (\n)
\psi_{\alpha '}(x)\vac+\psi^+_{\alpha '}(x)\epsilon (\n)\left\{
\psi_{\alpha '}(x), A^+_\alpha (\k)\right\}\vac= \nonumber \\
&&=-\bar{v}(x)\left\{\phi_\beta (x), A^+_\alpha (\k)\right\}\epsilon
(\n)v(x)\phi^+_\beta (x)\vac+ \nonumber \\
&&+u_0^2 \phi^+_{\alpha '} (x)\epsilon
(\n)\left\{\phi_{\alpha '} (x), A^+_\alpha (\k)\right\}\vac=   \label{26}\\
&&=-\bar{v}(x)\frac{g(\k)}{(2\pi)^\frac{3}{2}}e^{i\k\x}\epsilon (\n)v(x)
\phi^+_\alpha (x)\vac+u_0^2\frac{g(\k)}{(2\pi)^\frac{3}{2}}e^{i\k\x}
\epsilon (\k)\phi^+_\alpha (x)\vac \nonumber
\eea
The last term in (\ref{26}) when being integrated over the space variables
is proportional to the operator \Aas ,and gives therefore normal c-value
contribution into the energy spectrum. Rather complicated situation arises
after integrating the first term. Evidently, the diagonalization of this
term cannot be done with an arbitrary function $v(x)$. So, we need to find
the conditions under which the diagonalization is possible. Taking into
account that $v(x) = v_{0}e^{i\alpha (x)}$ we can write:  \be
\bar{v}(x)\epsilon (\n)v(x)\phi^+_\alpha (x) = v_0^2\epsilon (\n)
\phi^+_\alpha (x) + \bar{v}(x)\left[\epsilon (\n), v(x)\right]
\phi^+_\alpha (x)
\label{27}
\ee
Here square brackets mean commutator we need to calculate.

Let $f(x)$ be an arbitrary function , then one may write
$$
\frac{1}{v_0} \left[\epsilon (\n), v(x)\right]f(x) = \left[\epsilon (\n),
e^{i\alpha (x)}\right]f(x) = -\frac{1}{2m}\left[(\n)^2, e^{i\alpha (x)}\right]
f(x) =
$$
$$
= -\frac{1}{2m}\left( 2\n e^{i\alpha (x)}\n f(x) + f(x)(\n)^2 e^{i\alpha (x)}
\right)
$$
Further we have
$$
(\n)^2 e^{i\alpha (x)} = e^{i\alpha (x)}\left( i(\n)^2\alpha (x) -
(\n \alpha (x))^2\right) .
$$
therefore
\be
\left[ \epsilon (\n), v(x)\right] = \frac{v(x)}{2m}\left( 2(\n \alpha (x))
\hat{\k} - i\n^2 \alpha (x) + (\n \alpha (x))^2\right),
\label{28}
\ee
where $\hat{\k} = -i\n$.

 From the considerations above it can be seen that a diagonalization is
possible only at $\n \alpha (x) = const.$ In this case the coefficient at
\fas will be homogeneous and the space integration leads to the term
proportional to \Aas. Keeping in mind all this we take $v(x)$ in the form
\be
\alpha (x) = \k_0 \cdot \x
\label{29}
\ee
where $\k_{0}$ is some constant vector, and its physical meaning we will
explain below. For the commutator we finally get
\be
\left[\epsilon (\n), v(x)\right] = \frac{v(x)}{2m}\left( 2\k_0 \hat{\k} +
\k_0^2\right)
\label{30}
\ee
Now we are able to write out the kinetic term contribution into the energy
spectrum
\bea
\left[ \psi^+_{\alpha '}(x)\epsilon (\n)\psi_{\alpha '}(x), A^+_\alpha (\k)
\right]\vac = u^2_0\epsilon (\k)\frac{g(\k)}{(2\pi)_\frac{3}{2}}e^{i\k\x}
\phi^+_\alpha (x)\vac - \nonumber \\
- v_0^2\epsilon (\k)\frac{g(\k)}{(2\pi)_\frac{3}{2}}e^{i\k\x}\left(
\epsilon (\n) + \frac{1}{2m}(2\k_0\cdot\hat{\k} + \k^2_0)\right)\phi^+_\alpha
(x)\vac
\label{31}
\eea
Integrating this relation along the whole space and accounting that
\be
\left(\epsilon (\n) + \frac{1}{2m}(2\k_0\cdot\hat{\k} + \k^2_0)\right)
e^{-i\k\x} = \epsilon (\k - \k_0)e^{-i\k\x}
\label{32}
\ee
we have
\bea
&&\left[ \int d^3x \psi^+_{\alpha '}(x)\epsilon (\n)\psi_{\alpha '}(x),
A^+_\alpha (\k)\right]\vac =    \nonumber \\
&& = \mid g(\k)\mid^2\left( u_0^2 \epsilon (\k) - v_0^2 \epsilon (\k - \k_0)
\right) A^+_\alpha (\k)\vac \label{33}
\eea
Combining (\ref{33}) and (\ref{25}) we obtain the expression for the energy
spectrum of excitation
\be
E(\k) =\mid g(\k)\mid^2\left( u_0^2 \epsilon (\k) - v_0^2 \epsilon (\k - \k_0)
-\lambda\frac{v_0^2}{V^*}\right)
\label{34}
\ee
Thus, transformation (\ref{16}) under condition (\ref{29}) realizes,
indeed, the dynamical mapping in a sense that the fields \fa, entering the
mapping, describe the Hamiltonian (\ref{1}) eigenstates. However, all this
is true due to the fact that the state without excitations $\vac$ is the
Hamiltonian eigenstate as well. Let us dwell on this question.

\vskip 1cm
{\bf 3. Vacuum stability and conterterm.}
\vskip 2mm

Let us compute the action of the Hamiltonian on the state $\vac$ without
excitations
\be
H\vac = \int d^{3}x \left[\psi_\alpha^{+}(x)\epsilon(\n)\psi_\alpha(x) +
\frac{\lambda}{4} \chi^{+}(x)\chi(x) \right]\vac.
\label{35}
\ee
For the kinetic term we have
$$
\psi_\alpha^{+}(x)\epsilon(\n)\psi_\alpha(x)\vac =
$$
\vspace{-0.3cm}
\bea
&&=\epsilon_{\alpha \beta}
\left( u_0\phi^+_\alpha (x) +\bar{v}(x)\epsilon_{\alpha \beta '}\phi_{
\beta '} (x)\right)\epsilon (\n)v(x)\phi^+_\beta (x)\vac =  \\
&&=\!\!\bar{v}(x)\phi_\alpha (x)\epsilon (\n)v(x)\phi^+_\alpha (x)\vac +
u_0\epsilon_{\alpha \beta}\phi^+_\alpha (x)\epsilon (\n)v(x)\phi^+_\beta (x)
\vac . \nonumber
\label{36}
\eea
The corresponding expression for the interaction term brings the
following result
\bea
\chi^{+}(x)\chi(x)\vac = -v(x)\chi^+(x)\psi_\alpha (x)\phi^+_\alpha (x)&\vac&=
\nonumber \\
= -v(x)\chi^+(x)\left(u_0\phi_\alpha (x) + v(x)\epsilon_{\alpha \beta}
\phi^+_\beta (x)\right)\phi^+_\alpha (x)&\vac& = \label{37} \\
= \frac{2u_0v(x)}{V^*}\chi^+ (x)\vac - v^2(x)\epsilon_{\alpha \beta}
\chi^+ (x)\phi^+_\beta (x)\phi^+_\alpha (x)&\vac&. \nonumber
\eea
For the first term in this sum we have
\bea
\chi^+ (x)\vac = \epsilon_{\alpha \beta}\psi^{+}_{\beta}(x)
\psi^{+}_{\alpha}(x)&\vac& = \nonumber \\
= \epsilon_{\alpha \beta}\psi^{+}_{\beta}(x)\left(u_0
\phi^+_\alpha (x) + \bar{v}(x)\epsilon_{\alpha \beta '}\phi_{\beta '}(x)
\right)&\vac& = \nonumber \\
= u_0\epsilon_{\alpha \beta}\left(u_0\phi^+_\beta + \bar{v}(x)\epsilon_
{\beta \beta '} \phi_{\beta'}(x)\right)\phi^+_\alpha (x)&\vac& = \label{38}\\
=- u_0^2\epsilon_{\alpha \beta}\phi^+_\alpha (x)\phi^+_\beta (x)\vac -
\frac{2\bar{v}(x)u_0}{V^*}&\vac& \nonumber
\eea
For the second one we obtain
\bea
&&\epsilon_{\alpha \beta}\chi^+ (x)\phi^+_\beta (x)\phi^+_\alpha (x)\vac =
\epsilon_{\alpha \beta}\epsilon_{\alpha '\beta '}\psi^+_{\beta '}(x)\psi^+_{
\alpha '}(x)\phi^+_\beta (x)\phi^+_\alpha (x)\vac= \nonumber \\
&&=\epsilon_{\alpha \beta}\epsilon_{\alpha '\beta '}\psi^+_{\beta '}(x)
\left( u_0\phi^+_{\alpha '} (x)+\bar{v}(x)\epsilon_{\alpha '\alpha ''}
\phi_{\alpha ''} (x)\right)\phi^+_\beta (x)\phi^+_\alpha (x)\vac= \label{39}\\
&&=\bar{v}(x)\epsilon_{\alpha \beta}\psi^+_{\beta '}\phi_{\beta '} (x)
\phi^+_\beta (x)\phi^+_\alpha (x)\vac= \nonumber \\
&&=-4\frac{\bar{v}^2(x)}{V^{*2}}\vac - \frac{2}{V^*}u_0\bar{v}(x)\epsilon_{
\alpha \beta}\phi^+_\alpha (x)\phi^+_\beta (x)\vac. \nonumber
\eea
Putting together (\ref{37}), (\ref{38}) and (\ref{39}) we get
\be
\chi^+ (x)\chi (x)\vac =
4\frac{v_0^2}{V^{*2}}\vac + \frac{2}{V^*}u_0v(x)\epsilon_{
\alpha \beta}\phi^+_\alpha (x)\phi^+_\beta (x)\vac
\label{40}
\ee
Thus, for the action of the Hamiltonian density on the vacuum we would have
\bea
&&\left(\psi_\alpha^{+}(x)\epsilon(\n)\psi_\alpha(x) +
\frac{\lambda}{4} \chi^{+}(x)\chi(x) \right)\vac = \nonumber \\
&&=\bar{v}(x)\phi_\alpha (x)\left( \epsilon (\n) +
\frac{\lambda}{2V^*}\right)v(x)\phi^+_\alpha (x)\vac + \label{41} \\
&&+ u_0\epsilon_{\alpha \beta}\phi^+_\alpha (x)
\left(\epsilon (\n) + \frac{\lambda}{2V^*}\right)v(x)\phi^+_\beta (x)\vac
\nonumber
\eea
Integrating this relation over the whole space we finally derive
\be
H\vac = const_0\vac + \Delta H(2)\vac \label{42}
\ee
where
\be
const_0 = \int d^3x \bar{v}(x)\left\{ \phi_\alpha (x),
\left(\epsilon (\n) + \frac{\lambda}{2V^*}\right)v(x)\phi^+_\alpha (x)
\right\}
\label{43}
\ee
and $\Delta H(2)$ is two-particle contribution equal to
\be
\Delta H(2) = \int d^3x u_0\epsilon_{\alpha \beta} \phi^+_\alpha (x)
\left(\epsilon (\n) + \frac{\lambda}{2V^*}\right)v(x)\phi^+_\beta (x)
\label{44}
\ee
The braces in (\ref{43}) stand for the anticommutator. Computing now
$const_0$ by passing in (\ref{43}) to the momentum representation
and taking into account that $\epsilon (\k)$ is quadratic function of
the momentum we can see

\bea
const_0&=&\frac{2}{V^*}\int d^3x \bar{v}(x)\left(\epsilon(\n) +
\frac{<\k^2>}{2m} + \frac{\lambda}{2V^*}\right) = \nonumber \\
&=&\frac{2v_0^2}{V^*}\int d^3x e^{-i\k_0\x}\left(\epsilon(\n) +
\frac{<\k^2>}{2m} + \frac{\lambda}{2V^*}\right)e^{i\k_0\x} =  \label{45} \\
&=& 2v_0^2\frac{V}{V^*}\left(\epsilon(\k_0) +
\frac{<\k^2>}{2m} + \frac{\lambda}{2V^*}\right)\nonumber
\eea
Here $V$ - is space volume, and
\be
<\k^2> = \frac{\int d^3k \k^2\mid g(\k)\mid^2}{\int d^3k
\mid g(\k)\mid^2}
\label{46}
\ee
$\Delta H(2)$ being expressed in terms of the creation operators \Aas of
quasiparticles has the form
\bea
\Delta H(2)&=&u_0v_0\epsilon_{\alpha \beta}\int d^3x \phi^+_\alpha (x)
\left( \epsilon(\n) + \frac{\lambda}{2V^*}\right)
e^{i\k_0\x}\phi^+_\beta (x) = \nonumber \\
&=& \int d^3k D(\k)\epsilon_{\alpha \beta}A^+_\alpha(\frac{\k_0}{2}-\k)
A^+_\beta (\frac{\k_0}{2}+\k)
\label{47}
\eea
where
$$
D(\k) = u_0v_0\bar{g}(\k-\frac{\k_0}{2})\bar{g}(\k+\frac{\k_0}{2})
\left( \epsilon(\k-\frac{\k_0}{2}) + \frac{\lambda}{2V^*}\right)
$$
Thus, as it follows from (\ref{42}), the non-excited state $\vac$ is not
an eigenstate of the Hamiltonian (\ref{1}) and the $\Delta H(2)$ term is
the source of the nonstationarity of $\vac$. This term describes correlated
fermionic couple of excitations, moving with the momentum $\k_{0}$. The
physical meaning of the relation (\ref{42}) is that it points to the
existence of energy exchange between the couple and the system of
fermions.  In order to account this exchange one must input into initial
Hamiltonian the term describing the couple. So, in the relation (\ref{42}) we
transfer $\Delta H(2)$ to the left part, thus, redefining the Hamiltonian
and taking as a physical Hamiltonian the quantity $H_{P}$ equal to
\be
H_p = :H - \Delta H(2):
\label{48}
\ee
where the normal ordering  is referred to the Hiesenberg fields.
If
$$\Delta H(2) = :\Delta H(2): + const,
$$
then
\bea
H_p = H - \Delta H(2) + const. \hspace{5 mm} H_p\vac = H\vac&-&\Delta
H(2)\vac+const\vac, \nonumber \\
H_p\vac = W_0\vac, W_0 = const_0&+&const.
\label{49}
\eea
Consequently, after substitution of the conterterm $\Delta H(2)$ into the
Hamiltonian (\ref{1}) the non-excited state $\vac$ becomes stationary, with
the energy equal to $W_{0}$. Will this redefinition of the Hamiltonian
change the one-particle excitation spectrum $E(\k )$? To answer the
question let us write $\Delta H(2)$ in the following form:
\bea
\Delta H(2)&=&u_0\epsilon_{\alpha \beta}\int d^3x \phi^+_\alpha (x)
\left( \epsilon(\n) + \frac{\lambda}{2V^*}\right)
v(x)\phi^+_\beta (x) - \nonumber \\
&-&u_0\epsilon_{\alpha \beta}\int d^3x \bar{v}(x)\phi_\alpha (x)
\left( \epsilon(\n) + \frac{\lambda}{2V^*}\right) \phi_\beta (x)
\label{50}
\eea
The second term, we have added to the sum (\ref{50}), does not contribute
to the relation (\ref{42}), however, providing the hermiticity of $\Delta
H(2)$. The contribution of $\Delta H(2)$ to the energy spectrum is defined
by the commutator (\ref{19}). It reads:
\be
\left[ \Delta H(2), A^+_\alpha (\k)\right] \vac\sim A_\alpha\vac = 0
\label{51}
\ee
Thus, the addition of the conterterm $\Delta H(2)$ to the initial
Hamiltonian does not affect the spectrum.

It is worth noting that the Hamiltonian of $H_{P}$ type has been studied in
the papers (\cite{5}), where some exact results for a set of physical
quantities have been obtained. Our case has two essential differences.
Firstly, $\Delta H(2)$ is added to the initial Hamiltonian not with an
arbitrary distribution function $D(\k )$ but in the form "provoked" by the
Hamiltonian $H$ itself. Secondly, $\Delta H(2)$ describes correlated pairs
not of the Heisenberg fields, but of the excitations \Aas. We will show
below that just these excitations form a bound state which is the ground
state of the interacting fermions.

At first sight $\Delta H(2)$ seems to break global gauge invariance.
However, this is no true. The fact is that coefficient at $D(\k )$, connected
with the rotation parameters $u_{0}$ and $v(x)$, is expressed in terms of
abnormal vacuum averages of the Heisenberg fields. So, under the gauge
transformations of the fields these parameters are transformed as well by
the way by which $\Delta H(2)$ remains invariant. In other words,
condensate of correlated fermionic pairs restores the guage symmetry of
the Hamiltonian.  The constant entering (\ref{49}) and defining the
energy of vacuum state $\vac$ we will calculate later on, but now we
consider two particle excitations of fermionic system with the Hamiltonian
$H_{p}$.

\vskip 1cm
{\bf 4. Wave function and Bound State Spectrum of Fermionic
Excitations.}
\vskip 2mm

Let us find how the Hamiltonian $H_{p}$ acts on a state composed of two
excitations with momenta $\k$ and $\q$
\bea
H_pA^+_\alpha (\k)A^+_\beta (\q)\vac&=&\left( W_0 + E(\k) + E(\q)\right)
A^+_\alpha (\k)A^+_\beta (\q)\vac + \nonumber \\
&+&\left\{ \left[ H_p, A^+_\alpha (\k)\right], A^+_\beta\right\}\vac
\label{52}
\eea
If the excitations do not interact with each other, the  last term in the
sum (\ref{52}) should be equal to zero. Then this two-particle excitation
will be an eigenstate of the Hamiltonian $H_{P}$ with the energy equal to the
sum of energies of both excitations. Now we will demonstrate that this term
does not vanish, reducing instead to the none-diagonal two-particle form.
In this case, however, it will turn out that there exists a correlated
combination of excitations which is stationary and defines the bound states
of these excitations. The problem now comes to the computation of the
last term in (\ref{52}). Earlier above we have calculated the action of the
commutator in (\ref{52}) on the $\vac$ state defining, thus, the
energy spectrum $E(\k )$. It is necessary here to take into consideration
those terms of the commutator as well, that vanished previously on the state
$\vac$. All contributions to the commutator rise now from the kinetic term,
interaction and conterterm $\Delta H(2)$. We begin treatment with the
kinetic term:
\bea
&&\left[ \psi^+_{\gamma}(x)\epsilon (\n)\psi_{\gamma}(x), A^+_\alpha (\k)
\right] \nonumber \\
= &&\bar{v}(x)\frac{g(\k)}{(2\pi)^\frac{3}{2}}e^{i\k\x}\epsilon_{\alpha
\beta}\epsilon(\n)\psi_\beta (x)+u_0\frac{g(\k)}{(2\pi)^\frac{3}{2}}e^{i\k\x}
\epsilon(\k)\psi^+_\alpha (x)
\label{53}
\eea
 From which follows:
\bea
&&\left\{ \left[ \psi^+_{\gamma}(x) \epsilon (\n)\psi_{\gamma}(x),
A^+_\alpha (\k)\right] , A^+_\beta (\q)\right\} = \nonumber \\
&&=\epsilon_{\alpha \beta}u_0\bar{v}(x)\frac{g(\k)g(\q)}{(2\pi)^3}\left(
 \epsilon(\q) + \epsilon(\k)\right) e^{i(\k+\q)}
\label{54}
\eea
Let us note that relations (\ref{53}) and (\ref{54}) are absolute, i.e.
they are fulfilled on any state. The analogous contribution coming from
conterterm reads
\bea
\left[-\Delta H(2), A^+_\alpha (\k)\right]&=&u_0\bar{v}(x)\epsilon_{\gamma
\gamma '}\left[\phi_\gamma (x)\left(\epsilon(\n) + \frac{\lambda}{2V^*}
\right)\phi_{\gamma '} (x), A^+_\alpha (\k)\right] = \nonumber \\
&=&u_0\bar{v}(x)\frac{g(\k)}{(2\pi)^\frac{3}{2}}e^{i\k\x}\epsilon_{\beta
\alpha}\left(\epsilon(\k) + \frac{\lambda}{2V^*}\right)\phi_\beta (x) +
\label{55} \\
&+&u_0\bar{v}(x)\frac{g(\k)}{(2\pi)^\frac{3}{2}}e^{i\k\x}\epsilon_{\beta
\alpha}\left(\epsilon(\n) + \frac{\lambda}{2V^*}\right)
\phi_\beta (x) \nonumber
\eea
and
\bea
&&\left\{ \left[ -\Delta H(2), A^+_\alpha (\k)\right],
A^+_\beta (\q)\right\} =  \nonumber \\
&&=\epsilon_{\beta \alpha} u_0\bar{v}(x)\frac{g(\k)g(\q)}{(2\pi)^3}e^{i(\k+\q)
\x}
\left( \epsilon (\q) + \epsilon (\k) + \frac{\lambda}{V^*}\right) .
\label{56}
\eea
Thus, as it follows from the relations (\ref{54}) and (\ref{56}) the
contributions into anticommutator from the kinetic term and conterterm are
just "c"-numbers. These contributions, as we will see later, compensate
each other so that the only term proportional to the coupling constant
$\lambda$ is left. Now we will show that this term, in its turn, is
canceled by the contribution coming from the interaction term
$\frac{\lambda}{4}\left(\chi^{+}(x)\chi(x)\right)$ of $H_{P}$. As a result
the sought anticommutator acting to the vacuum $\vac$ will be proportional
to the two-fermionic state. Now the role of the conterterm $\Delta H(2)$
is completely determined, it annihilates two-particle component in the
$\vac$ state and vacuum component in the \Aas \Abs $\vac$ state. Let us
find out the contribution of the four-fermionic interaction. For the
Heisenberg fields entering the relation (\ref{21}) one can write:
\bea
\psi^+_\alpha (x)\chi (x) &=& u_0^3\epsilon_{\gamma \gamma '}\phi^+_\alpha
(x)\phi_\gamma (x)\phi_{\gamma '}(x) -\frac{2u_0^2}{V^*}v(x)\phi^+_\alpha (x)
+ \nonumber \\
&+& 2u_0^2v(x)\phi^+_\alpha (x)\phi^+_\beta (x) \phi_\beta (x) -
\frac{2}{V^*}u_0v_0^2\epsilon_{\alpha \beta}\phi_\beta (x) + \label{57} \\
&+&2u_0v_0^2\epsilon_{\alpha \beta}\phi_\beta (x)\phi^+_\gamma (x)\phi_\gamma
(x) + v_0^2v(x)\epsilon_{\alpha \beta}\epsilon_{\gamma \gamma '}\phi_\beta
(x) \phi^+_\gamma (x) \phi^+_{\gamma '} (x) \nonumber
\eea
It is
easy now to calculate the action of the corresponding anticommutator on to
the vacuum. Omitting the simple calculations, amounting to the commutation of
\Aas with fields (\ref{57}) we have:
\be
\left\{ \psi^+_\alpha (x)\chi (x),
A^+_\beta (\q)\right\}\vac =
v(x)\frac{g(\q)}{(2\pi)^\frac{3}{2}}e^{i\q\x}\epsilon_{\alpha \beta}
\epsilon_{\gamma \gamma '}\phi^+_\gamma (x) \phi^+_{\gamma '} (x)\vac
\label{58}
\ee
Here we have used the fact that
\be
\phi^+_\alpha \phi^+_\beta \vac = \frac{1}{2}\epsilon_{\alpha \beta}
\epsilon_{\gamma \gamma '}\phi^+_\gamma (x) \phi^+_{\gamma '} (x)\vac
\label{59}
\ee
Further,
\bea
\left\{ \chi^+ (x)\psi_\alpha (x), A^+_\beta (\q)\right\}&=&
\chi^+ (x)\left\{ A^+_\beta (\q), \psi_\alpha (x)\right\} +
\left[ A^+_\beta (\q), \chi^+ (x)\right] \psi_\alpha (x) \nonumber \\
\left\{ A^+_\beta (\q), \psi_\alpha (x)\right\}&=&u_0\frac{g(\q)}
{(2\pi)^\frac{3}{2}}e^{i\q\x}\delta_{\alpha \beta} \label{60}
\eea
Unification of these relations gives
\bea
& &\left\{ \left[ \frac{\lambda}{4}\chi^+ (x)\chi (x), A^+_\alpha (\k)\right] ,
A^+_\beta (\q)\right\} \vac = \nonumber \\
&=&\frac{\lambda}{V^*}u_0\bar{v}(x)\epsilon_{\alpha \beta}\frac{g(\k)g(\q)}
{(2\pi)^3}e^{i(\k+\q)\x}\vac+ \label{61} \\
&+&\frac{\lambda}{2}\epsilon_{\alpha \beta}\frac{g(\k)g(\q)}
{(2\pi)^3}e^{i(\k+\q)\x}\epsilon_{\gamma \gamma '}\phi^+_\gamma (x)
\phi^+_{\gamma '} (x)\vac \nonumber
\eea
for the interaction term. As it can be seen from this result the first
addendum in (\ref{61}) really compensates the last addendum in (\ref{57})
so that the anticommutator acting to the vacuum gives the only
two-fermionic state. Integration over the whole space yields:
\bea
H_p A^+_\alpha (\q_1)A^+_\beta (\q_2)\vac&=& \left( W_0 + E(\q_1) +
E(\q_2)\right) A^+_\alpha (\q_1)A^+_\beta (\q_2)\vac + \nonumber \\
&+&\lambda\frac{g(\q_1)g(\q_2)}{(2\pi)^3}\int d^3x
e^{i(\q_1+\q_2)\x}\phi^+_\alpha (x) \phi^+_\beta (x)\vac \label{62}
\eea
Here we
would like to note that the excitations of two fermions with a total spin $1$
(symmetrical over $\alpha$ and $\beta$) do not interact with each other,
because in this case the last term in the sum (\ref{62}) is equal to zero
identically and the energy of the state presents just the sum of free
energies of either excitation.
Let us pass to the variables ${\bf P} = \q_1 + \q_2$, $\k = \frac{1}{2}\left(
\q_1 - \q_2\right)$. Then using the expression for the fields
$\phi_{\alpha}(x)$ via operators \Aa, we can rewrite the equation (\ref{62})
in the following form:
\bea
&&H_p A^+_\alpha(\frac{{\bf P}}{2}-\k)A^+_\beta(\frac{{\bf P}}{2}+
\k)\vac= \nonumber\\
=&&\left(W_0+E(\frac{{\bf P}}{2}-\k)+E(\frac{{\bf P}}{2}+
\k)\right) A^+_\alpha (\frac{{\bf P}}{2} -\k)A^+_\beta
(\frac{{\bf P}}{2} + \k)\vac + \label{63} \\
&&+\frac{\lambda}{(2\pi)^3}g(\frac{{\bf P}}{2} -
\k)g(\frac{\bf P}{2} + \k)\int d^3q \overline{g}(\frac{\bf
P}{2} - \q)\overline{g}(\frac{\bf P}{2} + \q)
A^+_\alpha(\frac{\bf P}{2} - \q)A^+_\beta(\frac{\bf P}{2} + \q)\vac,
\nonumber
\eea
Define the wave packet $\hat{A}^+(\bf P)$ as:
\be
\hat{A}^+(\p) = \int G(\k,\p)\epsilon_{\alpha \beta}
A^+_\alpha(\frac{\p}{2} - \k)A^+_\beta(\frac{\p}{2} + \k)d^3k
\label{64}
\ee
and choose the wave function $G(\k,\p)$ in order to $\hat{A}^+(\p)\vac$
be an eigenstate of the Hamiltonian $H$:
\be
H_p\hat{A}^+(\bf P)\vac = \left( W_0 + \mu (\p)\right)\hat{A}^+(\p)\vac
\label{65}
\ee
After the simple calculations we obtain from (\ref{63}) and (\ref{65}):
\be
G(\k,\p) = \gamma_0\cdot\frac{\overline{g}(\frac{\p}{2} -
\k)\overline{g}(\frac{\p}{2} + \k)}{\mu(\p) - E(\frac{\p}{2} - \k) -
E(\frac{\p}{2} + \k)},  \label{66}
\ee
where
$$
\gamma_0 = \frac{\lambda}{(2\pi)^3}\int d^3k G(\k, \p)g(\frac{\p}
{2} - \k)g(\frac{\p}{2} + \k).
$$
This set of relations determines the wave function $G(\k, \p)$ and the energy
of the two excitations bound state $\mu(\p)$, with $\p$ being momentum of the
bound state as a whole, so that $\mu(\p =0)$ is the bound state mass. In
order to obtain the equation on $\mu(\p)$ let us substitute the manifest
form of the function $G(\k, \p)$ (\ref{66}) into the expression for
$\gamma_0$. We see that $\gamma_0$ is cancelled from the left and right parts
and we have the following equation:
\be
1 = \frac{\lambda}{(2\pi)^3}\int d^3k\frac{\mid g(\frac{\bf P}{2} -
\k)\mid^2\mid g(\frac{\bf P}{2} + \k)\mid^2}{\mu(\p) - E(\frac{\bf P}{2} -
\k) - E(\frac{\bf P}{2} + \k)}. \label{67}
\ee
This equation is the necessary and sufficient condition for existence
of two particle bound state, and it determines the bound state energy
$\mu(\p)$.  It is worth noting that this kind of condition always appears
when searching the solution of the BS equation\cite{Kalin}, what follows from
its homogeneity, and equation (\ref{67}) corresponds to the zero determinant
of the homogeneous system. We will analyse this equation below, but now let
us turn to the calculation of the BS amplitude $\Phi_{\alpha\beta}(x,y;\p)$.
It contains all the information about bound state. As has been pointed out in
Introduction, to find it one has to solve the BS equation, which kernel in
its turn is determined by the solution of the SD equation. However, if the
dynamical mapping is known, it is possible to calculate
$\Phi_{\alpha\beta}(x,y;\p)$ directly, since, by definition, it is a matrix
element between two physical states for the product of the Heisenberg fields:

\bea
\Phi_{\alpha\beta}(x,y;\p)&=&
\left<0\mid T \psi_\alpha(x)\psi_\beta(y)
\mid A(\p)\right>,\nonumber\\
\left.\mid A(\p)\right>&\equiv&\hat{A}^+(\p)\vac. \label{68}
\eea
We can construct four BS amplitudes from the fields $\psi_\alpha(x)$.
Introducing the variables $\R = \frac{\x+\y}{2},\;\;\r =\x-\y$ after
straightforward calculations we can write for these amplitudes:
\bea
\left<0\mid T \psi_\alpha^+(x)\psi_\beta(y)\mid A(\p)\right>&=&
u_0v_0e^{-i\k_0\x}\epsilon_{\alpha\alpha'}\delta_{\beta\beta'}G_{\alpha'
\beta'}(x,y;\p); \nonumber \\
\left<0\mid T \psi_\alpha(x)\psi_\beta^+(y)\mid A(\p)\right>&=&
u_0v_0e^{-i\k_0\y}\epsilon_{\beta\beta'}\delta_{\alpha\alpha'}
G_{\alpha'\beta'}(x,y;\p); \nonumber \\
\left<0\mid T \psi_\alpha(x)\psi_\beta(y)\mid A(\p)\right>&=&
u_0^2\delta_{\alpha\alpha'}\delta_{\beta\beta'}G_{\alpha'\beta'}(x,y;\p);
\label{69} \\
\left<0\mid T \psi_\alpha^+(x)\psi_\beta^+(y)\mid A(\p)\right>
&=&v_0^2e^{-2i\k_0\R}\epsilon_{\alpha\alpha'}\epsilon_{\beta\beta'}
G_{\alpha'\beta'}(x,y;\p), \nonumber
\eea
where the vertex part $G_{\alpha\beta}(x,y;\p)$ is defined by the matrix
element:
\bea
&&G_{\alpha\beta}(x,y;\p) =
\left<0\mid\phi_\alpha(x)\phi_\beta(y)\mid A(\p)\right> = \label{70}\\
&&= -\frac{2}{(2\pi)^3}e^{i\R\p}\epsilon_{\alpha\beta}\int d^3k
G(\k,\p)
g(\frac{\p}{2}+\k)g(\frac{\p}{2}-\k)e^{i\k\r}\dot e^{-iE(\frac{
\p}{2}+\k)t_x-iE(\frac{\p}{2}-\k)t_y}. \nonumber
\eea
Consider this relation at equal times $t_x=t_y=t$. We have:
\bea
&&G_{\alpha\beta}(x,y;\p)\mid_{t_x=t_y=t} =
-\frac{2\epsilon_{\alpha\beta}}{(2\pi)^{\frac{3}{2}}}e^{i\R\p-i\mu(\p)t}\dot
\gamma_0q(\r,t;\p), \label{71}\\
&&q(\r,t;\p)=\frac{1}{(2\pi)^{\frac{3}{2}}}\int d^3k \frac{
\mid g(\frac{\p}{2}+\k)\mid^2\mid g(\frac{\p}{2}-\k)\mid^2e^{i\k\r}}{
\mu(\p) - E(\frac{\p}{2}-\k) - E(\frac{\p}{2}+\k)}\dot e^{it\left(
\mu(\p) - E(\frac{\p}{2}-\k) - E(\frac{\p}{2}+\k)\right)}.\nonumber
\eea
The plain wave appeared in (\ref{71}) describes a movement of the bound state
as a whole with the momentum $\p$ and the energy $\mu(\p)$, and form factor
$q(\r,t;\p)$ is related with the bound state internal structure. At
$t=0,\;\;\p=0\;\; q(\r,0;0)\equiv \Phi(\r)$ coinsides with Schredinger wave
function in $"x"$ representation. Let us consider it in details:
\be
\Phi (\x) = \frac{1}{(2\pi)^\frac{3}{2}}\int d^3k\frac{\mid g(\k)
\mid^2}{\mu - E(\k) - E(-\k)}e^{i\k\x}
\label{72}
\ee
is thought to be the bound state wave function in the configuration space.
The bound state has the structure of the pair strictly correlated over
the momenta.  Parameter $\gamma_{0}$ is determined by the normalization of
distribution $G(\k,\p)$. Due to the fact that $\lambda \leq 0$ the following
relation
\be
\mu - E(\k) - E(-\k) < 0 \label{73}
\ee
should be satisfied for all $\k$, since only in this case the signs of the
left and right sides of the equation (\ref{67}) will be conformed and the
function under the integral will not have the peculiar points in the whole
integration region.

Let us calculate now the wave function ${\Phi}(\x )$, considering
the case when $\epsilon (\k ) = \frac{k^{2}}{2m}$ and representing $G(\k )$
in the following form:
\be
\frac{\mid g(\k)\mid^2}{E(\k) + E(-\k) - \mu} = \frac{M}{\k^2 + \delta^2}
\label{74}
\ee
Then the nonequality (\ref{73}) corresponds to the conditions:
\be
M > 0, \;\;\;\delta > 0
\label{75}
\ee
By the use of the manifest form of the energy spectrum (\ref{34}) we obtain:
\bea
\hspace{20 mm} M&=&\frac{m}{1-2v_0^2}>0 \nonumber \\
\hspace{20 mm} \delta^2&=&-\frac{1}{1-2v_0^2}\left( v_0^2\k_0^2 + \frac{
2m\lambda}{V^*}v_0^2 + \frac{m\mu}{\mid g(\k)\mid^2}\right) > 0.
\label{76}
\eea
The approximation can be made by putting $\mid g(\k )\mid^{2} = 1$ and the
integral (\ref{72}) will remain still convergent. As far as $\mid g(\k)
\mid^{2}$ plays the role of a cutting-off factor at the large momenta,
this approximation will lead to distortion of ${\Phi}(\x )$ only in the
region $\x \sim 0$. Having in mind all mentioned above we can see, that
from equation (\ref{76}) follows:
\bea
\mu <
-2v_0^2\left(\frac{\k_0^2}{2m}\right. &+&\left.  \frac{\lambda}{V^*}
\right) \simeq 2E(0) \nonumber \\ 1 - 2v_0^2 > 0 \label{77}
\eea
Evidently, $E(0)$ defines the lower boundary of the energy spectrum
$E(\k)$.  The relations (\ref{77}) represent the necessary conditions for
existing of the bound state of the fermionic excitations. There are two
possible different physical situations according to the sign of $E(0)$.
In the case when $E(0) \leq 0$ the state $\vac$ with the energy
$W_{0}$ is located in the excitation spectrum, and the bound state is the
lowest, being at least $2E(0)$ distant from the lower boundary of spectrum.
Because $E(\k )$ at $\k$ large enough becomes positive, there exists such a
momentum ${\bf p}_{0}$ for which $E({\bf p}_{0}) = 0$. Thus the state
\be
\mid 0,1> = a\vac + c_\alpha A^+_\alpha ({\bf p}_0)\vac,
\label{78}
\ee
where $a$ and $c_{\alpha}$ - any complex numbers, is a stationary state
with the energy $W_{0}$. If $E(0) > 0$, then for the value of $\mu$ three
variants are possible. Firstly, $E(0)\leq \mu < 2E(0)$. Then the
bound state lies within the excitation spectrum and the $\vac$ state with
the energy $W_{0}$ will be the ground state with the energy gap between it
and the excitation spectrum. Secondly, when $0\leq \mu \leq E(0)$ two
levels system arises with the ground state equal to $\vac$. And
lastly, if $\mu < 0$, we will have again two levels system, though the
bound state will be the ground state of two excitations with the energy
equal to $\mu$. In the particular case when $E(0) = 0$ the momentum
$\k_{0}$ is determined by the four-fermionic coupling constant $\lambda$.
 From the expression (\ref{77}) we obtain:
\be
\k_0^2 = - \frac{2m\lambda}{V^*}
\label{79}
\ee

For the wave function $\Phi (\x)$ we have, up to the normalization:
\be
\Phi (\x) = \frac{1}{(2\pi)^\frac{3}{2}}\int d^3k\frac{\mid g(\k)
\mid^2}{E(\k) + E(-\k) - \mu}e^{i\k\x} \simeq M\int d^3k \frac{e^{i\k\x}}
{\k^2+\delta^2}
\label{80}
\ee
Integrating it over we get:
\be
\Phi (\x) = M\frac{2\pi^2}{r}e^{-r\delta}
\label{81}
\ee
As a consequence of the approximation we see the singularity of (\ref{81})
at $r = 0$. In order to account correction to (\ref{81}) let us introduce
the screening parameter $r_{0}$:
\be
\Phi (\x)\Rightarrow \Phi (\x) = M\frac{2\pi^2}{r+r_0}e^{-r\delta}
\label{82}
\ee
 From here we obtain
$$
r_0 = M\frac{2\pi^2}{\Phi (0)}
$$
Further, as it follows from (\ref{80})
\bea
\Phi (0)&=&\frac{1}{(2\pi)^\frac{3}{2}}\int d^3k\frac{\mid g(\k)
\mid^2}{E(\k) + E(-\k) - \mu}\simeq \nonumber \\
&\simeq& \frac{1}{(2\pi)^\frac{3}{2}}
\int d^3k\frac{\mid g(\k)\mid^4}{E(\k) + E(-\k) - \mu} =-\frac{(2\pi)^3}{
\lambda}
\label{83}
\eea
where we have used the equation (\ref{67}). Finally we derive
\be
r_0 = -\frac{\lambda M}{4\pi}
\label{84}
\ee
It is seen from the solution (\ref{82}) that quantity $\frac{1}{2\delta}$
determines the radius of the bound state. Note, at last, that the
wave function ${\Phi}(\x )$ satisfies the equation
\be
\left( E(\n) + E(-\n) - \mu\right) \Phi (\x) = \frac{1}{2}(2\pi)^3\left\{
\phi_\alpha (\x), \phi^+_\alpha (0)\right\},
\label{85}
\ee
which follows from the manifest form of ${\Phi}(\x )$.

As a next step let us examine the solutions of the equation (\ref{67}).
Multiplication of the numerator and the denominator under the integral on
the same quantity $E(\k ) + E(-\k ) - \mu$ gives for $\mu$:
\be
\mu = 2\left< E(\k)\right>_G + \frac{(2\pi)^3}{\lambda\int d^3k G^2(\k)}
\label{86}
\ee
where
$$
2\left< E(\k)\right>_G = \frac{\int d^3k G^2(\k)\left( E(\k) + E(-\k)\right)}
{\int d^3k G^2(\k)}
$$
Substituting the explicit form of the spectrum (\ref{34}) into $\left<
E(\k ) \right>_{G}$ and supposing that $\mid g(\k )\mid^{2} \simeq 1$ we
obtain:
\be
\mu = 2\frac{\left< \k^2\right>}{2M} - v_0^2\frac{\k_0^2}{V^*} +
\frac{(2\pi)^3}{\lambda\int d^3k G^2(\k)}
\label{87}
\ee
Hereupon one can see that if the quantity $M$ is interpreted as an
effective mass of the excitation then the first term in (\ref{86}) represents
the sum of the kinetic energies each of the excitations. The last two terms
describe the bond energy of the excitations.

The solution of the equation (\ref{67}) we obtain as an asymptotic series
over the coupling constant $\lambda$. Using again the approximate expression
for $\delta^{2}$, viz. keeping $\mid g(\k )\mid^{2} \approx 1$ as in
(\ref{79}) we can write equation (\ref{67}) in the following form:
\be
1 =
-\frac{\lambda M}{(2\pi)^3}\int d^3 \frac{\mid g(\k)\mid^2}{\k^2+\delta^2}
\label{88}
\ee
This equation defines $\delta^{2}$ as a function of $\lambda$. We will seek
solution of (\ref{88}) in the form of the asymptotic series
\bea
\delta^2&=& a_0 + a\lambda +\frac{c_1}{\lambda} + \frac{c_2}{\lambda^2} +
\cdots \nonumber \\
\k^2 + \delta^2&=& \lambda \left( a + \frac{\k^2 a_0}{\lambda} + \frac{c_1}{
\lambda^2} + \frac{c_2}{\lambda^3} +\cdots \right)
\label{89}
\eea
By means of the series conversion formulae we can derive
\be
\frac{\lambda}{\k^2 + \delta^2} = \frac{1}{a+b\epsilon +c\epsilon^2 +\cdots}
= \frac{1}{a}\left[ 1 -\frac{b}{a}\epsilon +\left( \frac{b^2}{a^2}
- \frac{c}{a}\right) \epsilon^2 +\cdots \right]
\label{90}
\ee
where $\epsilon = \frac{1}{\lambda},\;\;b = a_{0} + \k^{2}, \;\; c = c_{1}$,
ect.  Substituting the series expansion (\ref{90}) into the equation
(\ref{88}), collecting the terms of the same power over $\lambda$ and
equating them to zero, we will obtain the chain of relations defining the
coefficients $a_{0},\; a,\; c_{1}$, ect. Straightforward calculations give
the following result:
\bea
&&a_0 =-\left<\k^2\right>, \hspace{5 mm} a = -\frac{M}{V*}, \nonumber \\
&&c_1 = -\frac{M}{V*}\frac{\int d^3k \mid g(\k)\mid^2\left(\k^2 -
\left<\k^2\right>\right)^2}{\int d^3k \mid g(\k)\mid^2}\equiv -\frac{V^*}{M}
\sigma \label{91} \\
&&\left<\k^2\right> = \frac{\int d^3k \k^2\mid g(\k)\mid^2}{\int d^3k
\mid g(\k)\mid^2} \nonumber
\eea
from which we receive the asymptotic series for $\delta^{2}$ and $\mu$:
\bea
\delta^2&=& -\left<\k^2\right> - \frac{M}{V^*}\lambda - \frac{V^*}{M}\cdot
\frac{\sigma}{\lambda} + \cdots, \nonumber \\
\mu &=& - \frac{\delta^2}{M} + 2E(0), \hspace{5 mm} 2E(0) = \left( \frac{
\k_0^2}{2m} + \frac{\lambda}{V^*}\right)\cdot\left( \frac{m}{M} - 1\right) .
\label{92}
\eea
According to the condition (\ref{76}) $\delta > 0$. This condition the
restricts region of the admissible values of $M$ and henceforth
$v_{0}^{2}$.  The magnitude of $M$ is determined from the vacuum energy
$W_{0}$ minimization condition.

 From the relations (\ref{91}) and (\ref{92}) it is easy to see that the
nonperturbative and singular contributions over the coupling constant
$\lambda$ into the energy are defined by the dispersion  $\sigma$ over
momentum distribution of $\mid g(\k )\mid^{2}$ inside the excitation.

\vskip 2mm
{\bf 5. Minimization of the Vacuum Energy $W_{0}$ With
Respect to the Transformation Parameters}.
\vskip 2mm

We  will go on now calculating the vacuum energy $W_{0}$ and minimizing it
with respect to the parameters. This procedure corresponds physically to
the fact that at the spontaneous transition, described by the
transformations (\ref{6}), the system itself fixes the magnitudes of
the parameters, choosing the states with the lowest energy.

As it follows from the relations (\ref{49}), the vacuum energy $W_{0}$
is equal to:
\be
W_0 = \frac{2v_0^2}{V^*}\int d^3x e^{-i\alpha(x)}\epsilon (\n)e^{i\alpha (x)}
+ 2v_0^2\frac{V}{V^*}\left(\frac{\left<\k^2\right>}{2m} + \frac{\lambda}{
2V^*}\right) + const.
\label{93}
\ee
where $ const$. is defined by result of bringing the conterterm $\Delta
H(2)$ to the normal ordering form with respect to the Heisenberg fields.
$v_{0}$ and $\alpha (x)$ are the transformation parameters. When we have
been obtaining the spectrum of one-particle excitations, as a necessary
condition of their existence we have found that $\alpha (x) = \k_{0}\x$ .
We retain $\alpha (x)$ in the expression for the vacuum energy $W_{0}$ as
an arbitrary function determining it later from the conditions of
the absolute minimum existing for $W_{0}$ over the variables $v_{0}$ and
$\alpha (x)$.  As it will be demonstrated in a
moment we will have the same result, what points to the strict correlation
between the existence of minimum and the one-particle excitation.

To calculate $ const$. in (\ref{93}) it is convenient to pass in the relation
(\ref{50}) from the fields $\phi_{\alpha}(x)$ to $\psi_{\alpha} (x)$
in accordance with the transformations (\ref{6}). As a result we obtain:
\bea
-\Delta H(2)&=& u_0\left( u_0^2 - v_0^2\right)\epsilon_{\alpha \beta}
\psi^+_\beta (x)\left(\epsilon (\n) + \frac{\lambda}{2V^*}\right) v(x)
\psi^+_\alpha (x) + \nonumber \\
&+& u_0\left( u_0^2 - v_0^2\right)\bar{v}(x)\epsilon_{\alpha \beta}
\psi_\alpha (x) \left(\epsilon (\n) + \frac{\lambda}{2V^*}\right)\psi_\beta
(x)- \nonumber \\
&-& 2u_0^2v_0^2\psi^+_\alpha (x)\left(\epsilon (\n) + \frac{\lambda}{2V^*}
\right)\psi_\alpha (x) + \label{94} \\
&+& 2u_0^2\bar{v}(x)\psi_\alpha (x)\left(\epsilon (\n) +
\frac{\lambda}{2V^*}\right) v(x)\psi^+_\alpha (x). \nonumber
\eea
We can see that the only term in (\ref{94}) which does not have the
normal ordering form is the last addendum. Therefore
\bea
const.&=& -2u_0^2\int d^3x\bar{v}(x)\left\{\psi_\alpha (x),
\left(\epsilon (\n) + \frac{\lambda}{2V^*}\right) v(x)\psi^+_\alpha (x)
\right\} = \label{95} \\
&=& -\frac{4u_0^4v_0^2}{V^*}
\int d^3x e^{-i\alpha(x)}\epsilon (\n)e^{i\alpha (x)} -
4\frac{V^*}{V}u_0^2v_0^2\left(\frac{\left<\k^2\right>}{2m} +
\frac{\lambda}{2V^*}\right). \nonumber
\eea
Together with the representation (\ref{93}) this leads to the following
expression for the energy $W_{0}$
$$
\frac{V^*}{V}W_0 = \rho (v_0)F(\alpha ) + \gamma (v_0)\Lambda,
$$
\vspace{-5 mm} where
\bea
F(\alpha )&=&\frac{1}{V}\int d^3x e^{-i\alpha(x)}\epsilon (\n)e^{i\alpha (x)}
, \hspace{5 mm} \Lambda = \frac{\left<\k^2\right>}{2m} +
\frac{\lambda}{2V^*}, \label{96} \\
\rho (v_0)&=& 2v_0^2\left( 1 -2(1-v_0^2)^2\right),\;\;\;\; \gamma (v_0) =
-2v_0^2(1 - 2v_0^2) \nonumber
\eea
The functional $F(\alpha )$ turns out to be essentially positive quantity.
Varying it with respect to $\alpha (x)$ and putting the first variation to
be equal to zero, it is easy to find that an extremum of $F(\alpha )$
(minimum in fact) is achieved on the functions satisfying the equation
$\n^{2}\alpha (x) = 0$. Evidently, the extremum would define a minimal
value of the energy $W_{0}$ provided that $\rho (v_{0}) > 0$. Therefore,
the minimization of $W_{0}$ over $v^{2}_{0}$ should be achieved within the
region where this condition is fulfilled.
 From the equation for $\alpha (x)$ one can easily write out the only
solution as $\alpha (\x ) = \k_{0}\x$. Let us study now the behavior of
$W_{0}$ as a function of parameter $v^{2} = -\kappa$. There are
three possible cases corresponding to the three different regions of
$\frac{\Lambda}{\epsilon (\k_{0})}: 1 +\frac{\Lambda}{\epsilon (\k_{0})} <
0;\; 1 +\frac{\Lambda}{\epsilon (\k_{0})} = 0;\; 1 +\frac{\Lambda}{\epsilon
(\k_{0})} > O$.

The value of $W_{0}$ at the minimum point should be negative and, besides,
in this point $\rho (\kappa_{0})>0$. It is seen from the explicit
form of $\rho (\kappa)$ (relation (\ref{96})) that this condition is
realized within to regions, when $\kappa = -v^{2}_{0} > 0$ and when
$-0.5 < \kappa < -1 + \frac{1}{\sqrt{2}}\simeq -0.3$. The left boundary  of
this inequality is determined by the condition (\ref{76}).
The equation on $\kappa_{0}$ is obtained from
the condition that the derivative of $W_{0}$ with respect to $\kappa$ is
equal to zero and has the form
\be
\kappa_0^2 + \frac{2}{3}\kappa_0\left( 2 + \frac{\Lambda}{\epsilon
(\k_{0})}\right) + \frac{1}{6}\left( 1 + \frac{\Lambda}{\epsilon
(\k_{0})}\right) = 0.  \label{97} \ee
Hence we have
\be
\kappa_0 =
-\frac{1}{3}\left( 2 + \frac{\Lambda}{\epsilon (\k_{0})}\right) +
\frac{1}{3}\left[\left( 2 + \frac{\Lambda}{\epsilon (\k_{0})}\right)^2 -
\frac{2}{3}\left( 1 + \frac{\Lambda}{\epsilon (\k_{0})}\right)
\right]^{\frac{1}{2}} \label{98}
\ee
At $1 + \frac{\Lambda}{\epsilon (\k_{0})} \rightarrow \infty,
\kappa_{0}\rightarrow -\frac{3}{8} \simeq -0.375$; and at $\kappa_{0} = -1 +
\frac{1}{\sqrt{2}}$ from (\ref{98}) follows that $1 + \frac{\Lambda}
{\epsilon (\k_{0})}\simeq 2$. Thus, the vacuum energy $W_{0}$ has the absolute
minimum with respect to the variables $\alpha (x)$ and $v_{0}$, provided
that $1 + \frac{\Lambda}{\epsilon (\k_{0})} > 2$, and position of this
minimum lies within the region $0.3 < v_{0}^{2} < 0.375$. The excitation
mass $M = \frac{m}{1-2v_{0}^{2}}$ is larger then the fermionic
mass $m$. There are no solutions inside the region $0 \leq 1 +
\frac{\Lambda}{\epsilon (\k_{0})} < 2$. Finally, if $1 + \frac{\Lambda}
{\epsilon (\k_{0})} < 0$ then minimum $W_{0}$ is situated at the region of
negative values of $v_{0}^{2}$, and transformation (\ref{6}) corresponds to
the hyperbolic rotation. Using expression for $\Lambda$ and (\ref{96}) we
get
\be
\frac{\k_0^2}{2m} + \left(\frac{\left<\k^2\right>}{2m} +
\frac{\lambda}{2V^*}\right) < 0
\label{99}
\ee
Let us rewrite this relation on the form either
\be
\frac{\k_0^2}{2m} + \frac{\lambda}{V^*} +
\left(\frac{\left<\k^2\right>}{2m} -
\frac{\lambda}{2V^*}\right) < 0
\label{100}
\ee
or
\be
\frac{\k_0^2}{2m} + \frac{\lambda}{V^*} < \left(\frac{\left<\k^2\right>}{2m}
+ \frac{\lambda}{2V^*}\right) < 0
\label{101}
\ee
Hence it follows that $2E(0) < 0$. Thus, the state $\hat{A}^{+}\vac$ will
be the lowest state with the energy equal to $\mu$. The excitation mass $M$
in this case will be less than $m$. Further, contribution to the bound
state energy $\mu$ corresponding to the kinetic energy of the pair moving
as a whole (second term in the sum (\ref{87})) will have positive sign but
the third term, describing interaction of the excitations - negative sign,
in agreement with the physical picture.

\vspace{1cm}
{\bf Conclusion.}

Finally we want to emphasize some aspects of this paper. The basic and
most essential one is postulating the dynamical mapping, but the
transformations (\ref{5}) and (\ref{6}) have been used just as a motivation
sample to find the mapping (\ref{16}).

The next important step is connected with the introduction of the
distribution function $g(\k)$, which allows to define the product of the
physical fields at the same space-time point. This fact corresponds to the
"cut-off" at large momenta, and the free parameter $V^*$ (spatial volume of
excitation), appearing in our model, is related with the "cut-off" parameter
$\Lambda_{cut}$ by the expression:
\vspace{-5mm}
$$
V^* = \frac{3}{4\pi}\left(\frac{2\pi}{\Lambda_{cut}}\right)^3.
$$
Such a smooth way of regularization leads to the fact that the
anticommutation relations for $\phi_\alpha(x)$ (\ref{10}) cease to be local,
that results, in its turn, in the arising of abnormal nonzero
anticommutators between the Heisenberg fields.

As a nontrivial cosequence of $"g"$-regularization we have the fact that in
all dynamical characteristics - energy spectrum, vacuum energy density,
bound state mass,- the coupling constant $\lambda$ is renormalized, such that
the effective coupling constant turns out to be $\frac{\lambda}{V^*}$.

The method of dynamical mappings in the form used here can be easily
generalized on the relativistic model (NJL), at least in separable
approximation. This problem will be considered in the forthcoming paper.

\begin{flushleft}
\underline{Aknowlegments}
\end{flushleft}
The authors are grateful to Profs. D.V.Shirkov, V.N.Pervushin,
Yu.L.Kalinovsky and V.A.Naumov for the crucial comments, useful disscusions
and support.

\begin{center}
{\bf Appendix.}
\end{center}

Here we present the formulas for some physical quantities.

I. Condensates.
\vskip -5mm
\begin{eqnarray*}
\left<0\mid\psi_\alpha(x)\psi_\beta^+(x)\mid 0\right>&=&
\frac{u_0^2}{V^*}\delta_{\alpha\beta}; \\
\left<0\mid\psi_\alpha^+(x)\psi_\beta(x)\mid 0\right>&=&
\frac{v_0^2}{V^*}\delta_{\alpha\beta};\\
\left<0\mid\psi_\alpha^+(x)\psi_\beta^+(x)\mid 0\right>&=&
\frac{u_0v_0}{V^*}\epsilon_{\alpha\beta}e^{-i\k_0\x};\\
\left<0\mid\psi_\alpha(x)\psi_\beta(x)\mid 0\right>&=&
-\frac{u_0v_0}{V^*}\epsilon_{\alpha\beta}e^{i\k_0\x}.
\end{eqnarray*}
\vspace{-3.5cm}
\begin{flushright}
(A.1)
\end{flushright}
\newpage

II. Casual Green functions for the Heisenberg fields.

\begin{eqnarray*}
\left<0\mid T \psi_\alpha(x)\psi_\beta^+(y) \mid 0\right>&=&
u_0^2G_{\alpha\beta}^{(1)}(x-y)+v_0^2e^{i\k_0(\x-\y)}
G_{\alpha\beta}^{(2)}(x-y);\\
\left<0\mid T \psi_\alpha^+(x)\psi_\beta(y)\mid 0\right>&=&
v_0^2e^{-i\k_0(\x-\y)}G_{\alpha\beta}^{(1)}(x-y)+
u_0^2G_{\alpha\beta}^{(2)}(x-y);\\
\left<0\mid T \psi_\alpha^+(x)\psi_\beta^+(y)\mid 0\right>&=&
u_0v_0\epsilon_{\alpha\beta'}\left(e^{-i\k_0\x}G_{\beta'\beta}^{(1)}(x-y)-
e^{-i\k_0\y}G_{\beta'\beta}^{(2)}(x-y)\right);\\
\left<0\mid T \psi_\alpha(x)\psi_\beta(y)\mid 0\right>&=&
-u_0v_0\epsilon_{\alpha\beta'}\left(e^{i\k_0\x}G_{\beta'\beta}^{(1)}(x-y)-
e^{i\k_0\y}G_{\beta'\beta}^{(2)}(x-y)\right),
\end{eqnarray*}
\vspace{-3cm}
\begin{flushright}
(A.2)
\end{flushright}
\vspace{1.5cm}

where $G_{\alpha\beta}^{(i)}(x-y),\;\;i=1,2$ are the casual Green functions
for the physical fields:
\begin{eqnarray*}
G_{\alpha\beta}^{(1)}(x-y)&=&
\left<0\mid T \phi_\alpha(x)\phi_\beta^+(y)\mid 0\right>;\\
G_{\alpha\beta}^{(2)}(x-y)&=&
\left<0\mid T \phi_\alpha^+(x)\phi_\beta(y)\mid 0\right>.
\end{eqnarray*}
\vspace{-2cm}
\begin{flushright}
(A.3)
\end{flushright}
\vspace{1cm}

III. Superpropagators.

Fourier image of the function $G_{\alpha\beta}^{(i)}(x-y)$, by
the conventional terminology (see e.g.,\cite{VolPer}) defines the
superpro\-pagator presented by the infinite seria of free field propagators
with initial spectrum $\epsilon(\k)$. Let us find this representation. We
have, for example:
\[
G_{\alpha\beta}^{(1)}(x-y)=\frac{i\delta_{\alpha\beta}}{(2\pi)^4}\int
\frac{\mid g(\k)\mid^2}{\omega -E(\k) + i\epsilon }e^{i\k(\x-\y)-i\omega(
t_x-t_y)}d^3kd\omega.
\]
\vspace{-1.5cm}
\begin{flushright}
(A.4)
\end{flushright}

Extracting the initial spectrum $\epsilon(\k)$ from the spectrum $E(\k)$:
\begin{eqnarray*}
E(\k)&=&\mid g(\k)\mid^2\left(u_0^2 \epsilon
(\k)-v_0^2 \epsilon (\k-\k_0)-\lambda\frac{v_0^2}{V^*}\right)\simeq\\
&\simeq& u_0^2\epsilon (\k)-v_0^2 \epsilon (\k-\k_0)-\lambda\frac{v_0^2}{V^*}=
\epsilon(\k)-\Sigma(\k),
\end{eqnarray*}
\vspace{-2cm}
\begin{flushright}
(A.5)
\end{flushright}

where $\Sigma(\k) = v_0^2\left(\epsilon(\k)+\epsilon(\k-\k_0)+
\frac{\lambda}{V^*}\right)$,
we come to the expression for the superpropagator:
\begin{eqnarray*}
\frac{1}{\omega -E(\k) + i\epsilon}&=& \sum_{m=1}^\infty(-1)^{m-1}\dot
 ( \Sigma(\k))^{m-1}\dot\Delta^m(\k), \\
\Delta(\k)&=& \frac{1}{\omega - \epsilon(\k) + i\epsilon}
\end{eqnarray*}
\vspace{-2cm}
\begin{flushright}
(A.6)
\end{flushright}

\vspace{0.5cm}

We would like to mention that self-energy part of $\Sigma(\k)$
is not an analytical function over the four fermionic coupling constant, this
fact is related with non-analyticity of $v_0^2(\lambda)$.

\end{document}